\documentclass[preprint,12pt]{elsarticle}

\usepackage{amssymb}

\usepackage{graphicx}
\usepackage{url}

\usepackage{graphicx}
\usepackage{algorithm}
\usepackage{algorithmic}
\usepackage{epstopdf}
\usepackage{listings}
\usepackage{amsfonts}
\usepackage{array}
\journal{Journal of Applied Logic}
\begin{document}

\begin{frontmatter}
%
%%% Title, authors and addresses
%
%%% use the tnoteref command within \title for footnotes;
%%% use the tnotetext command for theassociated footnote;
%%% use the fnref command within \author or \address for footnotes;
%%% use the fntext command for theassociated footnote;
%%% use the corref command within \author for corresponding author footnotes;
%%% use the cortext command for theassociated footnote;
%%% use the ead command for the email address,
%%% and the form \ead[url] for the home page:
%%% \title{Title\tnoteref{label1}}
%%% \tnotetext[label1]{}
%%% \author{Name\corref{cor1}\fnref{label2}}
%%% \ead{email address}
%%% \ead[url]{home page}
%%% \fntext[label2]{}
%%% \cortext[cor1]{}
%%% \address{Address\fnref{label3}}
%%% \fntext[label3]{}
%
%\title{A Novel Methodology for Early Soft Error and Mitigation Analysis for FPGA-based Space Applications}
\title{Formal Analysis of SEU Mitigation for Early Dependability and Performability Analysis of FPGA-based Space Applications}
%% use optional labels to link authors explicitly to addresses:
% \author[label1,label2]{}
% \address[label1]{}
% \address[label2]{}

\author[label1]{Khaza Anuarul Hoque\corref{mycorrespondingauthor}}
\cortext[mycorrespondingauthor]{Corresponding author}
\ead{khaza.hoque@cs.ox.ac.uk}
\author[label1]{Otmane Ait Mohamed}
\author[label2]{Yvon Savaria}

\address[label1]{Concordia University, Montreal, Canada}
\address[label2]{Polytechnique Montr\'eal, Montreal, Canada}

\begin{abstract}
SRAM-based FPGAs are increasingly popular in the aerospace industry due to their field programmability and low cost. However, they suffer from cosmic radiation induced Single Event Upsets (SEUs). In safety-critical applications, the dependability of the design is a prime concern since failures may have catastrophic consequences. An early analysis of the relationship between dependability metrics, performability-area trade-off, and different mitigation techniques for such applications can reduce the design effort while increasing the design confidence. This paper introduces a novel methodology based on probabilistic model checking, for the analysis of the reliability, availability, safety and performance-area tradeoffs of safety-critical systems for early design decisions. Starting from the high-level description of a system, a Markov reward model is constructed from the Control Data Flow Graph (CDFG) and a component characterization library targeting FPGAs. The proposed model and exhaustive analysis capture all the failure states (based on the fault detection coverage) and repairs possible in the system. We present quantitative results based on an FIR filter circuit to illustrate the applicability of the proposed approach and to demonstrate that a wide range of useful dependability and performability properties can be analyzed using the proposed methodology. The modeling results show the relationship between different mitigation techniques and fault detection coverage, exposing their direct impact on the design for early decisions.

\end{abstract}

\begin{keyword}

Probabilistic model checking \sep FPGA, Dependability \sep Performability \sep Markov Reward Model \sep SEU \sep CDFG
%% keywords here, in the form: keyword \sep keyword

%% PACS codes here, in the form: \PACS code \sep code

%% MSC codes here, in the form: \MSC code \sep code
%% or \MSC[2008] code \sep code (2000 is the default)

\end{keyword}

\end{frontmatter}
\section{Introduction}
\subsection{Motivation}

With respect to power consumption and speed, reconfigurable computing with Field Programmable Gate Arrays (FPGAs) can outperform general-purpose CPUs, and due to their field programmability, absence of non-recurring engineering costs, low manufacturing costs and other advantages, SRAM-based FPGAs are increasingly attractive compared to Application-Specific Integrated Circuits (ASICs). Unfortunately, a great disadvantage of these devices is their sensitivity to radiation effects which can cause bit flips in memory elements and ionisation induced transient faults in semiconductors, commonly known as Single Event Upsets (SEUs) \cite{XilinxRosetta,NASAScrub}. Different vendors have provided radiation hardened FPGAs to meet the requirements of the avionic and space industries \cite{Xilinx5QV}. However, these devices are very expensive as they are manufactured in relatively low volumes and they also lag by two or three technology nodes (28-nm\footnote{nm = nanometer} CMOS\footnote{CMOS = Complementary Metal–-Oxide-–Semiconductor}, 45-nm CMOS, 65-nm CMOS etc.) when compared to commercial products. Therefore, there is a growing need to analyze the possible utilization of commercial SRAM-based FPGA components in harsh radioactive environments such as outer space.

To deal with SEUs, designers mostly rely on redundancy-based solutions, such as Triple Modular Redundancy (TMR) \cite{XilinxTMR} for high reliability, and \emph{configuration memory (Configuration Bits) scrubbing} \cite{NASAScrub} in order to mitigate SEUs and to gain high availability. Scrubbing is traditionally done in the order of milliseconds, however such fast scrubbing consumes high power \cite{ScrubPower,scrub_power_thesis2_dynamic} and hence scrubbing at a lower frequency is desired \cite{scrub_power_thesis}. The strict power budgets typical of deep space missions such as Voyager-1, Voyager-2 \cite{voyager}, and even the Mars missions have produced a need for delayed scrubs (in the order of hours or days) to save power. Scrubbing is often used in conjunction with other forms of mitigation techniques such as TMR or spare components, to increase reliability. However, in cases where performability (reliability and performance combined) is a major concern, redundancy-based solutions might not always be the default choice \cite{khaza}. Much of the literature also reported approaches for safety modeling, and dependability improvement mostly based on improving the fault detection coverage \cite{cov1,cov2,cov3}. Unfortunately the relationship between fault detection coverage, scrub interval, redundancy, rescheduling \cite{rescheduling,CDFGFault}, performability-area trade-off, and how these impact each other in the early design option evaluation was ignored. Always setting a target of 100\% fault detection coverage is expensive in terms of time and cost and is unnecessary in many cases. That is why, to choose the right design options and parameters, it is important to evaluate the relationship between the reliability, availability, safety, performability with the adopted fault mitigation technique, fault detection coverage and mission time. Such analysis at an early design stage will allow the designer to develop more reliable and efficient solutions, and may also reduce the overall cost associated with the design effort. Our work aims to achieve these goals.

\subsection{Contributions and Limitations of Previously Reported Approaches}
This paper proposes a methodology that can be applied at early design stages to evaluate various design options of reconfigurable systems in terms of dependability and performability-area tradeoff. The proposed methodology is based on probabilistic model checking \cite{PMC1stochastic}. The main advantage of using probabilistic model checking is the exhaustive nature of the analysis, which results in numerically exact answers to temporal logic queries \cite{advancesPMC}. This contrasts with discrete-event simulations in which approximate results are generated by averaging results from a large number of random samples.

Using our approach, for each design option, a Markov chain dependability model is constructed from the Control Data Flow Graph (CDFG) representation of the system under analysis. The Markov chain dependability model captures all the possible components failures considering their fault detection coverage parameters, and possible recovery by rescheduling and scrubbing. Each state of the model is then augmented with associated performance and area rewards obtained using the high-level synthesis technique. The cumulative reward of this single Markov Reward Model (MRM) \cite{MRM2} is then used to evaluate the corresponding design option in terms of  reliability, availability, safety and performability-area trade-off. Current work in this area \cite{intro-related, soap, Related_DFG_MRM,cov1} either separates the dependability analysis from performance/area, coverage analysis, or does not analyze such safety-critical applications at an early design stage. Commercial tools for dependability analysis, such as \emph{isograph} \cite{website:ISOGraph}, concentrate mainly on the reliability and availability analysis and lack the support for performability evaluation. Since the PRISM probabilistic model checker \cite{PMC2prism} allows reward modeling, our work overcomes this limitation.

Our previously proposed modeling method was limited to only reliability, availability and performability-area trade-off evaluation \cite{khaza}, hence it was not possible to reason about the system's safety, or the relationship between the fault detection coverage with fault mitigation parameters. In this work, we extend our previous model by capturing the concept of safety using the notion of fault detection coverage. The quantitative results from our obtained model show  some important observations such as the fact that high fault detection coverage is not always helpful for gaining high reliability, and that scrubbing delay also has a considerable impact. In terms of safety, we also show how the scrubbing interval affects the safety of available design options with the same fault detection coverage.

Our analysis also shows that redundancy may fail to improve reliability if it has lower fault detection coverage compared to a design with no redundancy but high fault detection coverage for some cases. For performability-area trade-off analysis, in our previous work, we showed that redundancy-based solutions might not always be the best choice as one may expect. Alternatively, for those cases, \emph{rescheduling} in conjunction with \emph{scrubbing} can be a good option. In this paper, we observe that if the scrub interval is small, our conclusion holds \cite{khaza}, even for a lower fault detection coverage. On the other hand, for a longer scrub interval, the design options show a different trend while we vary the fault detection coverage. To our knowledge, this is the first attempt to evaluate such relationships at early design stages using probabilistic model checking.

The remainder of the paper is organized as follows. Section 2 introduces some basic concepts about dependability metrics, the background of SEU effects, SEU mitigation techniques and probabilistic model checking. Section 3 reviews CDFG rescheduling and related works. The proposed methodology and modeling details are discussed in section 4, and in section 5, we present quantitative results from an FIR filter case study illustrating our proposed methodology. Section 6 concludes the paper with future research directions.

\section{Preliminaries}

\subsection{Single Event Upsets}
FPGAs are configurable logic devices that implement logic circuits with a fabric that includes Look-up tables (LUTs), memories and routing resources that connect the LUTs and memories. In a reconfigurable FPGA, the configuration memory is a collection of bits commonly known as a bitstream. Bitstream bits set the values of the LUT, flip-flop and memory initialization values, and states of switches and connection boxes that route signals through the FPGA. For Virtex devices from Xilinx, the configuration memory is composed of SRAM cells. Since the FPGA configuration bitstream is stored in volatile SRAMs, interaction with high-energy radiated particles that are common in the aerospace environment, such as protons, neutrons, and heavy ions, may corrupt the FPGA configuration. The effects of these particles on electronics are collectively known as Single-Event Effects (SEE) and there are several types of SEE that are relevant to FPGAs. Single Event Upsets (SEUs) occur when one or more bits in configuration memory changes state due to a radiation event. The state of the FPGA configuration memory defines the architecture of the application. As a consequence, SEUs in the configuration memory are not only harmful but also could result in the catastrophic failure of the design.
\subsection{Scrubbing}
Data scrubbing is a well-known technique for error correction. It uses a background task that periodically inspects memory for errors and corrects the errors using Error-Correcting Code (ECC) memory or the protected duplicate copy of data. Scrubbing in FPGAs such as Xilinx uses a similar approach for scrubbing the configuration memory. FPGA scrubbing is divided into different categories based on the type of implementation (external vs internal), and based also on the scrubbing strategy (blind vs read-back) \cite{intvsExtScrub}.  A scrubbing strategy is composed of at least one correction technique and, optionally, a  detection technique. External blind scrubbing, which is modeled in this paper, is a very popular and reliable scrubbing strategy that requires no additional detection algorithm before fixing the configuration-memory upsets. Correction techniques for external blind scrub usually use off-chip rad-hard memory to store the golden file (also known as the golden data or golden bitstream - which is a copy of the initial configuration memory loaded onto the device) and then periodically (or continuously) reconfigures the FPGA with that golden bitstream to repair the SEUs. The blind scrubbing strategy is very popular in FPGA-based space platforms because of its effectiveness (they can fix any number of upsets) and simplicity (less implementation complexity). Since frequent scrubbing drains power, for many power-aware space applications delayed scrubbing is a good choice \cite{Khaza_MEMOCODE2014}. Scrubbing can be performed at a specified rate, meaning that there might be a period of time between the moment the upset occurs and the moment when it is repaired. That is why scrubbing is usually implemented with another form of mitigation, such as a redundancy-based solution known as TMR~\cite{XilinxTMR}.

\subsection{Probabilistic Model Checking}
Model checking~\cite{Clarke86automaticverification} is a well-established technique used in both industry and academia to verify the correctness of finite-state concurrent systems. In contrast, \emph{Probabilistic model checking} deals with systems that exhibit stochastic behaviour, and  is based on the construction and analysis of a probabilistic model of the system, typically a Markov chain. In this paper, we focus on the Continuous-time Markov Chains~(CTMCs) and Markov reward models~\cite{MRM2} widely used for reliability and performance analysis.\\

\noindent \textbf{Definition 1.} \emph{A (labeled) CTMC $\mathcal{M}$ is a tuple (S, \textbf{R}, L) with S as a finite set of states,  $ \textbf{R}: S \times S \rightarrow \mathbb{R}_{\geq 0}$ is the rate matrix, and $ L : \longrightarrow 2^{AP}~$  is the labeling function that assigns every state $s \in S$ a set L(s) of atomic propositions $a \in AP$ which are valid in s.}\\

\noindent  The rate \textbf{R}$(s,s')$ defines the delay before which a transition between the states $s$ and $s'$ takes place. Intuitively, $\textbf{R}(s,s') \geq 0$ iff there is a transition from $s$ to $s'$. Furthermore, $1 - e^{-{\small{\textbf {R}}}(s,s')\times t} $  is the probability  that the transition $s \longrightarrow s'$ can be triggered within  $t$ time unit. Such exponentially distributed delays are  suitable for modelling component lifetimes and inter-arrival times. If $\textbf{R}(s,s') \geq 0$ for more than one state $s'$, then it initiates a competition between these transitions originating in $s$, commonly known as the race condition.\\

\noindent \textbf{Remark 1.} We allow self-loops in our CTMC model, and according to Definition 1, self-loops at state $s$ are possible and are modeled by having $\textbf{R}(s,s) > 0$. The inclusion of self-loops neither alters the transient nor the steady-state behavior of the CTMC, but allows the usual interpretation of Linear-Time Temporal (LTL) operators (we refer the interested reader to \cite{pnueli1977temporal} for more details about the syntax and semantics of LTL) like the \emph{next step} ($\mathcal{X}$) that we will exploit in Section~5 to check the correctness of the model.\\

%A CTMC comprises a set of states $S$ and a transition rate matrix $ \textbf{R}: S \times S \rightarrow \mathbb{R}_{\geq 0}$. The rate \textbf{R}$(s,s')$ defines the delay before which a transition between states $s$ and $s'$ takes place. If \textbf{R}$(s,s') \ne 0$ then the probability that a transition between the states $s$ and $s'$ might take place within time $t$ can be defined as $1 - e^{-{\small{\textbf {R}}}(s,s')\times t} $. No transitions will take place if  \textbf{R}$(s,s') = 0$. Exponentially distributed delays are  suitable for modelling component lifetimes and inter-arrival times.

In the Probabilistic model checking approach using CTMCs, properties are usually expressed in some form of extended temporal logic such as Continuous Stochastic Logic (CSL), a stochastic variant of the well-known Computational Tree Logic (CTL) \cite{Clarke86automaticverification}.  \\

\noindent \textbf{Definition 2.} A CSL formula $\Phi$ defined over a CTMC $\mathcal{M}$ is one of the form:
\begin{center}
$\Phi ~::=~ true ~|~ a ~|~ \Phi \land \Phi ~|~ \neg \Phi ~|~ \mathcal{S}_{\bowtie p}(\Phi) ~|~ \mathcal{P}_{\bowtie p}(\phi)$\\
$\phi ~::=~ \mathcal{X} \Phi ~|~ \Phi \mathcal{U} \Phi ~|~ \Phi \mathcal{U}^{\leq t} \Phi$ \\
\end{center}

\noindent \emph{where $a \in AP$ is an atomic propositions, $p \in [0,1] $, $t \in \mathbb{R}_{> 0}$ and $ \bowtie \in \{<, \leq, \geq, > \}$. Each $\Phi$ is known as a state formula and each $\phi$ is known as a path formula.}\\

The detailed syntax and semantics of CSL can be found in \cite{Baier99approximatesymbolic}. In CSL, $\mathcal{S}_{\bowtie ~p} (\Phi)$ asserts that the steady-state probability for a $\Phi$ state meets the boundary condition $\bowtie~p$. On the other hand, $\mathcal{P}_{\bowtie~p} (\phi)$ asserts that the probability measure of the paths satisfying $\phi$ meets the bound given by $\bowtie ~p$. The meaning of the temporal operator $\mathcal{U}$ and $\mathcal{X}$ is standard (same as in LTL). The temporal operator $\mathcal{U}^{\leq t}$ is the real-time invariant of $\mathcal{U}$. Temporal operators like \emph{always} ($\Box$), \emph{eventually} ($\Diamond$) and their real-time variants ($\Box^{\leq t}$ and $\Diamond^{\leq t}$) can also be derived from the CSL semantics.  Below, we show some illustrative examples with their natural language translations:\\

%(\textbf{G} and \textbf{F} operator in PRISM respectively)

%In CSL, the $\mathcal{P}$ operator is used to reason about the probability of satisfying a formula (can be either a \emph{state} or \emph{path} formula), and the $\mathcal{S}$ operator is used to reason about the steady-state behaviour.   Below, we show some illustrative examples with their natural language translations:\\

%In CSL, given that $p \in [0,1]$ a real number, $ \unlhd \in {\leq, <, \geq, >}$, $\phi$ and $\Phi$ is a state and path formula respectively, $\mathcal{S}_{\unlhd ~p} (\phi)$ asserts that the steady-state probability for a $\phi$ state meets the boundary condition $\unlhd ~p$. On the other hand, $\mathcal{P}_{\unlhd ~p} (\Phi)$ asserts that the probability measure of the paths satisfying $\Phi$ meets the bound given by $\unlhd ~p$.
\noindent 1. $failure \Rightarrow P_{\geq 0.95} [ \neg ~fail ~U^{\leq 200} ~up]$  - ``Once a failure has occurred, with probability 0.95 or greater, the system will successfully recover within 200 hours and without any further failures occurring".\\
\noindent 2. $ \mathcal{P}_{\geq 0.98} [\Diamond ~complete]$ - ``The probability of the system eventually completing its execution successfully is at least 0.98". \\
\noindent 3. $\mathcal{S}_{\leq 10^{-9}} [Failure] $ - ``In the long run, the probability that a failure condition can occur is less than or equal to $10^{-9}$ ". \\

%\noindent 2. $failure \Rightarrow P_{\geq 0.95} [ \neg ~fail ~U^{\leq 200} ~up]$  - ``Once a 'failure' has occurred, with probability 0.95 or greater, the system will successfully recover (will be in \emph{up} mode) within 200 hours and without any further failures occurring".\\

\noindent In the PRISM property specification language \texttt{P, S, G, F, X} and \texttt{U} operators are used to refer to the $\mathcal{P, S, \Box, \Diamond, X}$ and $\mathcal{U}$ operator. In addition, PRISM also supports the expression $\mathcal{P = ? [\phi]}$ and $\mathcal{S = ?} [\Phi]$ in order to compute  the actual probability of the formula  $\phi$ and $\Phi$ being satisfied. Additional properties can be specified by adding the notion of rewards (the $\mathcal{R}$ operator) to CSL \cite{kwiatkowska2006model}. Each state (and/or transition) of the model is assigned a real-valued reward, allowing queries such as:\\

\noindent $\mathcal{R}_{= ?} [\Diamond ~success] $ - ``What is the expected reward accumulated before the system successfully terminates?". \\

\noindent Rewards can be used to specify a wide range of measures of interest, for example, the number of correctly delivered packets or the time that the system is operational. Of course, conversely, the rewards can be considered as costs, such as power consumption, expected number of failures, etc. PRISM also allows the use of customized properties using the \texttt{filter} operator: $filter (op, prop, states)$, where \emph{op} represents the filter operator (such as forall, print, min, max, etc.), \emph{prop} represents the PRISM property and \emph{states} (optional) represents the set of states over which to apply the filter.

%Model checking~\cite{Clarke86automaticverification} is a well established formal verification technique to verify the correctness of finite-state systems. Given a formal model of the system to be verified in terms of labelled state transitions and the properties to be verified in terms of temporal logic, the model checking algorithm exhaustively and automatically explores all the possible states in a system to verify if the property is satisfiable or not \cite{clarke1999model}. If not, a counterexample is generated. \emph{Probabilistic model checking} deals with systems that exhibit stochastic behaviour, such as fault-tolerant systems. Probabilistic model checking is  based on the construction and analysis of a probabilistic model of the system, typically a Markov chain. In this paper, we focus on the continuous-time Markov chains~(CTMCs) and Markov reward models~\cite{MRM2}, widely used for reliability and performance analysis.

%Model checking~\cite{Clarke86automaticverification} is a well established formal verification technique to verify the correctness of finite-state systems. \emph{Probabilistic model checking} deals with systems that exhibit stochastic behaviour, such as fault-tolerant systems. Probabilistic model checking is  based on the construction and analysis of a probabilistic model of the system, typically a Markov chain. In this paper, we focus on the continuous-time Markov chains~(CTMCs) and Markov reward models~\cite{MRM2}, widely used for reliability and performance analysis.
%

\section{Rescheduling-based Fault Recovery and Related Works}
\subsection{CDFG Rescheduling}
\begin{figure}
\centering
\includegraphics[width=0.8\textwidth]{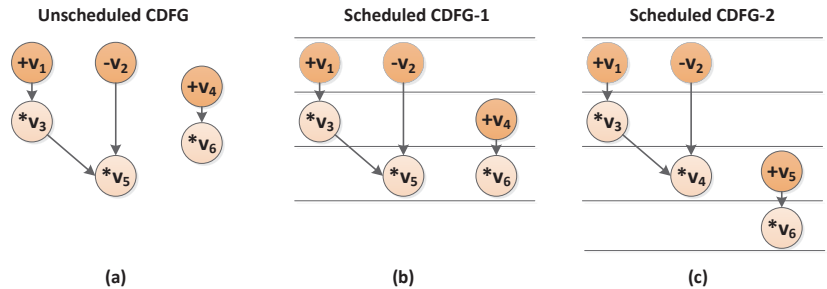}
\caption{CDFGs scheduled over available resources}\label{fig:CDFGall}
\end{figure}

Let us consider the CDFG of a synchronous dataflow DSP application shown in \figurename~\ref{fig:CDFGall}(a). Based on data dependencies, this application can be carried out in a minimum of three control steps ($c_{steps}$) using the CDFG-1 shown in \figurename~\ref{fig:CDFGall}(b), with two adders and two multipliers. Such an implementation provides a throughput of $1/3=0.33$ (for non-pipelined systems, throughput is the inverse of latency \cite{latency1, latency2}, throughput modeling will be addressed later in this paper). Another alternative consists of implementing the application with only one multiplier and two adders but in four control steps, as shown by CDFG-2 in \figurename~\ref{fig:CDFGall}(c). In this case the throughput is 0.25. Considering the priority of the throughput or area metric, the appropriate CDFG can be selected. Based on this idea, when a resource fails (due to a configuration bit flip), an alternative schedule can be derived to continue the system operation using the remaining resources, most likely at a lower throughput. For instance, to maximize the throughput, let us consider that CDFG-1 is implemented. For a single component failure, e.g. a multiplier, the application can be rescheduled to implement CDFG-2 with a lower throughput. Such a rescheduling-based fault tolerance approach was introduced in \cite{rescheduling,CDFGFault,FIR} for fault-secure microarchitectures and multiprocessors (a computation on a set of processors is fault-secure if no fault in the computation generated by a faulty processor goes undetected). For FPGA-based designs, such a fault recovery technique can be adopted as well and we explore the dependability and performability-area tradeoffs for such systems. It is of interest that the controller for rescheduling the operations is assumed to be fault-free. This controller can be implemented in a separate chip with proper fault-tolerance mechanisms.

\subsection{Related Works}

%\begin{figure}
%\centering
%\includegraphics[width=80mm]{Fig/CDFG1and2new.eps}
%\caption{CDFGs scheduled over available resources}\label{fig:CDFG-1}
%\end{figure}

For many years, dependability analysis of complex safety-critical systems has been an active research area in both academia and industry. Hence, researchers have put a lot of efforts
in analyzing fault tolerant systems. In \cite{TosunHLS}, the authors proposed a reliability-centric high-level synthesis approach to address SEUs. Their framework uses reliability characterization to select the most reliable implementation for each operation fulfilling latency and area constraints. In addition, researchers have dedicated considerable effort to modeling the behavior of gracefully degradable large-scale systems using continuous-time Markov reward models \cite{related_MRM1,related_MRM2}. In \cite{ana1}, the authors present the modeling and analysis of fault trees based on stochastic logic. To produce the models, probabilistic analysis of all different types of gates is carried out first, and then the probability models are converted to their equivalent stochastic logic gates. Compared to Markov chains, a classical fault tree is limited for modeling only non-repairable systems. The impact of fault detection coverage on reliability with quantitative assessment on different types of systems were performed and reported in \cite{covrel-1,covrel-2,covrel-3}. However, the relationship between different fault mitigation approaches for early design analysis was not explored in any of these works. In \cite{related_casestudy}, a case study is presented to measure the performance of a multiprocessor system using a continuous-time Markov reward model. An approach for analyzing the performance, area and reliability metric of a design using a Markov reward model is presented in \cite{Related_DFG_MRM}. The authors used transistor lifetimes to model the reliability and performance, hence the model is composed of non-repairable modules. The use of a non-formal commercial tool makes their approach quite rigid in terms of analysis. Moreover, in their proposed approach, the reward calculation is manual, as the traditional commercial tools for reliability analysis do not support reward modeling.

Even though our model has some similarities to their work \cite{Related_DFG_MRM}, our approach is more flexible because we use probabilistic model checking. Our work focuses on a different fault model: cosmic radiation-induced configuration bit-flips in FPGAs. Since scrubbing is possible in FPGA designs, we also add repair to our Markov reward model. In terms of the failure type, repair capability, inclusion of fault detection coverage, and use of a characterization library to model the system, the application of our work and our methodology is different and novel when compared to all the related works described above.

\section{Proposed Methodology}
\begin{figure}
\centering
\includegraphics[width = 0.7\textwidth]{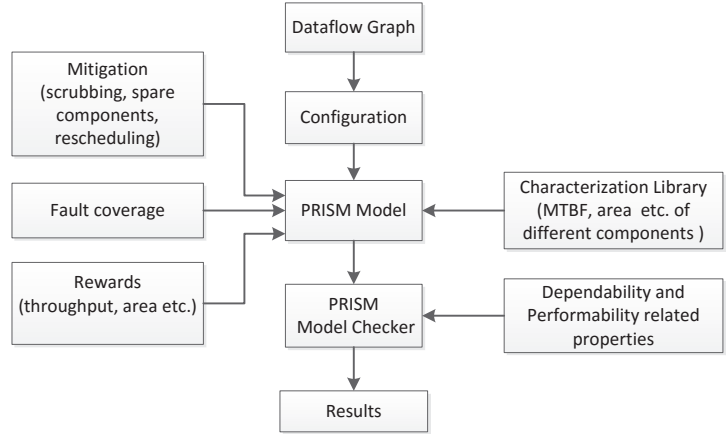}
\caption{Proposed methodology}\label{method}
\end{figure}

In \figurename~\ref{method}, we present the proposed methodology that starts from the dataflow graph of the application. The boxes in the methodology represent the steps, and the edges show the relationship between them. The steps are as follows:\\

\noindent 1. \emph{Dataflow graph:} The CDFG is extracted from a high-level design description expressed in C++. The idea is inspired from \cite{LibraryBasedSER}, however, we use a different tool known as GAUT \cite{GAUT} for this purpose.\\

\noindent 2. \emph{Configuration:} As mentioned earlier, a CDFG can be implemented with  different component allocations (design options). To analyze each configuration, we model them with the PRISM modeling language. From now on, we will refer to the term \emph{design options} as \emph{configurations} in the rest of the paper.\\

\noindent 3. \emph{PRISM modeling:} PRISM modeling requires the description of a given system in terms of component failure rates, adopted fault mitigation strategy, fault detection coverage and performance measures. To acquire the component failure rate, we use a \emph{characterization library} (characterization library is explained in section 4.4). The modeled fault mitigation techniques are: rescheduling, cold spare components and blind scrubbing. For rescheduling a CDFG, with available components if possible, a high-level synthesis algorithm, such as \emph{forced-directed list scheduling} \cite{Paulin} can be used. Since the model is parametric, the fault detection coverage and the scrub interval can be varied for analysis. Each state of this Markov model can be augmented with associated rewards such as throughput (obtained using high-level synthesis techniques: CDFG scheduling with available components in each state), area (measured in terms of the total number of LUTs required to implement the design, obtained from the component characterization library) or any other metric of interest. The resulting Markov Reward Model is then analyzed using the PRISM model checker tool.\\

\noindent 4. \emph{PRISM model checker:} The PRISM tool then computes the set of all states which are reachable from the initial state and identifies any deadlock states (i.e. reachable states with no outgoing transitions). PRISM then parses one or more temporal logic properties (e.g. in CSL) and performs model checking, determining whether the model satisfies each property  or compute the actual probability.\\

\subsection{Markov Modeling of Reliability and Availability}

CTMC models are very commonly used for modeling the dependability of gracefully degradable systems. Each state in a CTMC model representing a specific configuration can be classified into different types depending on the number of healthy components. For instance, the FIR filter in \figurename~\ref{fig_fir} (quantitative results section) requires at a minimum an adder and a multiplier for successful operation. Hence, any state that does not fulfill the minimum resource availability is labeled as a \emph{failed state}. At the end, the state labeled as \emph{all fail} represents a state in which all the components in the system have failed one-by-one due to SEUs. Note that \emph{safe} and \emph{unsafe} failures are not considered at this stage of modeling. How to include safety in the model will be described in detail in the next subsection. The initial state of a configuration has the maximum throughput and all the components are functional. The assumptions for our modeling are defined as follows:\\

%\begin{figure}
%\centering
%\includegraphics[width = 0.5\textwidth]{Fig/methodologyNew1.eps}
%\caption{Proposed methodology}\label{method}
%\end{figure}

\noindent \emph{Assumption 1}:~All the components fail independently and the time-to-failure for a component due to a configuration bit flip is exponentially distributed. Exponential distributions are commonly used to model the reliability of systems where the failure rate is constant. The \emph{scrub} interval is assumed to follow an exponential distribution as well \cite{li2013reliability1, li2013reliability2, li2013reliability3}, with a rate, $\mu$ = $1/\tau$, where $\tau$ represents the scrub interval. A scrub process that follows a deterministic time delay can be modeled using the Erlang process \cite{Erlang, Khaza_MEMOCODE2014, Hoque_DATE2015, hoque2016early} which is considered as a part of our future works \cite{hoque2016applying}.\\

\noindent \emph{Assumption 2}:~ Every component in the system is connected with the other components (via multiplexers). This assumption is needed for the simplicity of the hardware model. The control unit can be designed as a finite state machine implemented either as a hardwired or microcoded controller. Since in many systems datapath components dominate the area of the design compared to control units, these components can be much more vulnerable to SEU than control units. Hence, we only consider the failures of the datapath components in this work, and the modeling of control units is left for future works.\\

\noindent \emph{Assumption 3}:~$Cold~spare$ components are used to provide redundancy and are active only when the same type of component fails. The $cold~spare$ components are only error-prone due to cosmic radiations when they are active.\\

\noindent \emph{Assumption 4}:~The reconfiguration and rescheduling times (i.e. the time taken for the system to reschedule when a component fails and the time taken for repair via  scrubbing) are extremely small compared to the times between failures and repairs. The time required for rescheduling is at most a few clock cycles and the time required for scrubbing is only a few milliseconds.\\

\noindent \emph{Assumption 5}:~All the states in the CTMC model can be classified into three types: $operational$, where all the components are functional and the system has the highest throughput; $degraded$, where at least one of the components is faulty; and $failed$, where the number of remaining non-faulty components is not sufficient to perform successful operation and hence has a throughput of $0$. In PRISM, a \emph{formula} can be used to classify such states as shown in the PRISM code.

Our model is described as a number of modules in PRISM, each of which corresponds to a component of the system. Each module has a set of finite-ranged variables representing different types of resources. The domain of the variables represents the number of available components of a specific resource. The whole model is constructed as the parallel composition of these modules. The behaviour of an individual module is specified by a set of guarded commands in the following form:
\begin{center}
 \texttt{[act] <guard> $\rightarrow$ <rate> : <update>} ;
\end{center}

\noindent where \texttt{act} is an (optional) action label and the \texttt{guard} is a predicate over the variables of all the modules in the model. A \texttt{rate} is an expression which evaluates to a positive real number and the \texttt{update} is of the form:

\begin{center}
 \((z_1'=v_1\)) \& \((z_2'=v_2\)) \& ..... \& \(( z_n'=v_n \)) ;
\end{center}

\noindent where $z_1, z_2,  ...., z_n$ are the local variables of the module and $v_1, v_2, .... v_n$ are expressions over all variable of the model.  The interpretation of the command is that if the \texttt{guard} is satisfied, then the module can make the corresponding transition with that associated \texttt{rate}. A very simple command for a module with only one variable $z$ might be:

\begin{center}
 [ ] \((z = 0\)) $\rightarrow$ 7.5 : \((z' = z + 1\)) ;
\end{center}

\noindent which states that if $z$ is equal to $0$, then it will be incremented by one and this action occurs with a rate of $7.5$. A second more significant example, is an application that implements a function using 2 adders and 2 multipliers but requires at least 1 adder and 1 multiplier (in the case of failure due to SEU) for successful operation. Such a configuration in the PRISM modeling language can be described as shown in \figurename~3. Once each module is specified in such a manner, the PRISM model checker then performs a parallel composition of all the modules to build the complete Markov chain of the system specified. To model interactions between multiple modules, i.e. simultaneous changes in their state, we use \emph{synchronisation}, which is specified by augmenting guarded commands with \emph{action} labels. The rate of a synchronous (combined) transition is defined as the product of the rates for each command.

%
%\begin{figure}
%\centering
%\includegraphics[width=85mm]{Fig/CTMC_2A2M_new.pdf}
%\caption{Sample CTMC for reliability/availability analysis}\label{CTMC}
%\end{figure}

%\begin{center}
% \texttt{[] <guard> $\rightarrow$ <rate> : <action>} ;
%\end{center}

\begin{figure}[!t]
\small{
\lstset{
           basicstyle=\footnotesize\ttfamily,
           keywordstyle=\ttfamily,
           stringstyle=\ttfamily,
           commentstyle=\ttfamily,
          breaklines=true,
                    }
%\begin{lstlisting}

\begin{lstlisting}[numberstyle=\small, numbersep=0pt,frame=single]  % Start your code-block

module adder
 a : [0..num_A] init num_A;
 [] (a > 0) -> a*lambda_A : (a' = a - 1);
 [rep] (a <= num_A) -> miu : (a' = num_A);
endmodule

module mult
 m : [0..num_M] init num_M;
 [] (m > 0) -> m*lambda_M : (m' = m - 1);
 [rep] (m <= num_M) -> 1 : (m' = num_M);
endmodule

formula fail = (a =0)|(m =0);
formula oper = (a=num_A)&(m=num_M);
formula degrade = !fail & !oper;
\end{lstlisting}
}
\normalsize{\caption{PRISM modeling for a system with 2-adders and 2-multipliers}} \label{code}
\end{figure}

\begin{figure}[!t]
\centering
\includegraphics[width=85mm]{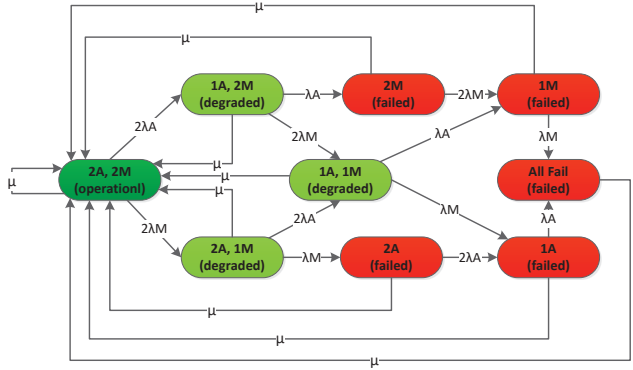}
\caption{Sample CTMC for reliability/availability analysis}\label{CTMC}
\end{figure}

\noindent In the PRISM code shown in \figurename~3, \texttt{num\_A} and \texttt{num\_M} represent the number of adders and multipliers available in the initial state of the configuration. The \texttt{lambda\_A} and the \texttt{lambda\_M} variables represent the associated failure rates of the adders and multipliers, whereas \texttt{miu} represents the repair rate. Each repair transition (scrub) leads back to the initial state, reflecting the scenario in which the configuration bit flips have been repaired. The value of \texttt{lambda\_A} and \texttt{lambda\_M} is obtained from a component characterization library, which will be explained later in the paper. PRISM then constructs, from this, the corresponding probabilistic model, in this case a CTMC. The resulting CTMC for this configuration is shown in \figurename~\ref{CTMC} (\texttt{lambda\_A}, \texttt{lambda\_M}, \texttt{miu}, \texttt{a} and \texttt{m} are reflected in the figure as $\lambda A, \lambda M$, $\mu$, $A$ and $M$ respectively). The repair commands in the code use the action label \texttt{[rep]} to synchronise the repair transitions between \texttt{module} \emph{adder} and \texttt{module} \emph{mult}. This demonstrates a phenomenon where, when the FPGA is scrubbed, all the components get fixed simultaneously. The intended repair rate is fully specified in \texttt{module} \emph{adder} and in \texttt{module} \emph{mult} specified as 1. As mentioned earlier, this is due to the fact that the rate of the synchronised transition is the product of the rates for each command. The \texttt{formula} \emph{fail}, \emph{oper} and \emph{degrade} classifies \emph{failed}, \emph{operational} and \emph{degraded} states in the model.

\subsection{Safety Modeling using Fault Detection Coverage}

From the reliability point of view, all failures are equal. However, from the perspective of \emph{safety}, failures can be divided into \emph{safe} and \emph{unsafe} ones.  \\

\noindent \textbf{Definition 3}: The safety \emph{S(t)} of a system at time \emph{t} is the probability that the system either performs its function correctly, or discontinues its operation in a fail-safe manner in the interval [\emph{0,t}] given that the system was operating correctly at time 0.\\

Any fault detection algorithm can be assumed to detect and handle all the faults properly; however, in reality this is not the case. A fault can escape the implemented fault detection mechanism. The fault detection coverage (or simply fault coverage\footnote{\emph{fault coverage} and \emph{coverage} both refer to the term \emph{fault detection coverage} throughout this paper.}) of a component can be defined by a conditional probability \emph{C}, that given the existence of a fault, the system detects it \cite{dubrova2013fault} :

\begin{center}
\noindent $C = P (fault~detection|fault~existence)$
\end{center}

For instance, a typical industrial requirement is that $99\%$ of single stuck-at faults are detected during manufacturing tests of ASICs \cite{rennels1984fault}. The fault detection coverage $C=0.99$ (a perfect fault detection coverage refers to C = 1) can be used as the measure of the system's ability to meet such a requirement which can be validated using simulation or emulation based fault injection techniques \cite{stott2008fault, stott2008fault1}. It is worth mentioning that with increasing fault detection coverage requirement, the cost of test development and test application also increases rapidly. Hence, it is necessary to analyze the relationship between the dependability metrics and the fault detection coverage at early design stages. This will enable the designers to set the target fault detection coverage for the design to be implemented based on the dependability requirements instead of aiming at an unnecessary higher coverage value.

In our case, if a fault escapes the detection mechanism then the system will not be able to reschedule, hence the system will continue its operation in a faulty mode. This means that each component in the configuration that implements the CDFG can fail either in a \emph{safe} or in an \emph{unsafe} fashion. This is why we need to refine the model by taking the fault detection coverage into account while introducing the concept of the \emph{safe failure} and \emph{unsafe failure}. We define them as follows:\\

\noindent \textbf{Definition 4}: \emph{Safe failure} is when a component fails due to an SEU, which is detected and handled by rescheduling depending on the number of remaining components. If the number of remaining components are not sufficient for rescheduling, the system moves to a \emph{fail safe} state.\\

\noindent \textbf{Definition 5}: An \emph{Unsafe failure} is defined as the fail silent behavior, e.g. when a system fails to detect a component's failure.\\

\noindent If all the faults are safely detected, it will eventually lead to the \emph{failed safe} state, whereas even if there is a single \emph{Unsafe failure} occurrence, it will immediately lead to the \emph{failed unsafe} state.\\
\begin{figure}
\centering
\includegraphics[height=45mm]{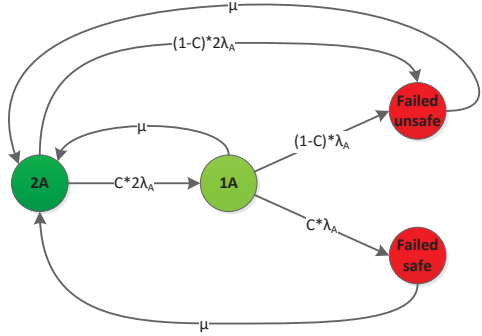}
\caption{Safety modeling of simple system with safe and unsafe failure}\label{coverage}
\end{figure}

\noindent \figurename~\ref{coverage} shows the modeling of safety for a simple single component system with only two adders including the repair transitions. For this case, we assume that the system requires at least one adder for a successful add operation. Initially the system is in the operational mode with two adders. When one adder fails, if the failure is detected, the system is rescheduled and continues with only one adder. This \emph{Safe failure} is modeled using the transition from state \emph{2A} to state \emph{1A} with a rate of $C*2\lambda_A$, where $C$ refers to the fault detection coverage and $\lambda_A$ refers to the failure rate of an adder. If the failure is not detected, then it moves to the \emph{Failed unsafe} state and this situation is modeled using the transition from state \emph{2A} to state \emph{Failed unsafe} with a rate of $(1-C)*2\lambda_A$. If another adder fails, the system will not be able to continue its operations, hence it will fail safely leading to the \emph{Failed safe} state. However, if this failure is not detected, then the system will eventually fail in an unsafe fashion. Inclusion of safety in the model requires the modification of assumption 5 as follows: \\

\noindent \emph{Assumption 5}:~All the states in the CTMC model can be classified into four types:\\

\noindent1. \emph{operational} - All the components are functional and the system has the highest throughput. \\
2. \emph{degraded} - At least one of the components is faulty. \\
3. \emph{failed~safe} - The number of remaining non-faulty components is not sufficient to perform successful operation and hence has a throughput of $0$. To reach the \emph{failed~safe} state, all the failures leading to this state must be \emph{Safe failures}.  \\
4. \emph{failed~unsafe} - At least one failure is not detected by the detection algorithm.  \emph{Unsafe failure} of a component immediately leads to the \emph{failed~unsafe} state.

\begin{figure}[!t]

\small{
\lstset{
           basicstyle=\footnotesize\ttfamily,
           keywordstyle=\ttfamily,
           stringstyle=\ttfamily,
           commentstyle=\ttfamily,
          breaklines=true,
          }
%\begin{lstlisting}

\begin{lstlisting}[numberstyle=\small, numbersep=0pt,frame=single]  % Start your code-block

module adder
a : [0..num_A+1] init num_A;
[] (a  > 0 & (a < num_A+1)) -> c*a*lambda_A :
(a'=a-1) + a*(1-c)*lambda_A : (a'= num_A+1);
[rep] (a >= 0 )   -> repair : (a'=num_A);
endmodule

module mult
m : [0..num_M+1] init num_M;
[] (m  >  0 & (m < num_M+1)) -> c*m*lambda_M :
(m'=m-1) + m*(1-c)*lambda_M : (m'= num_M+1);
[rep] (m >= 0) -> 1 : (m'=num_M);
endmodule

formula fail_unsafe=((a=num_A+1)|(m=num_M+1));
formula fail_safe=((a=0)|(m=0))& !fail_unsafe;
formula oper =(a=num_A)&(m=num_M);
formula degrade=!fail_safe&!fail_unsafe&!oper;

\end{lstlisting}
}
\normalsize{\caption{PRISM modeling refined after inclusion of coverage \emph{(c)} for a system with 2-adders and 2-multipliers}} \label{fig:Markov_two_partition}
\end{figure}

\figurename~6 shows the modified PRISM code from \figurename~3 after including the fault detection coverage variable \emph{c} in the model.

\subsection{Peformability Modeling using MRM}

When a system changes its state from one to another one due to a full/partial failure or repair, the performance level can change. Such a scenario can be described by different states using a Markov model that provides a framework for combined performance-reliability (performability) analysis. Formally, an MRM consists of a CTMC $ X = X(t), t > 0$ with finite states space $S$, and a reward function $r$ where $ r: S \rightarrow \mathbb{R}$ \cite{performance_MRM}. For each state $ i \in S$, $r_i $ denotes the reward obtained per unit time spent by \emph{X } in that state which represents the performance level given by the system while it is in that state.

Performability measures can be distinguished in different classes, mainly into two: steady-state performability and transient or point performability. For $i \in S$, let $w_i$ denote the steady-state probability of residing in state $i$, and $p_i(t)$ the (transient) probability of residing in state $i$ at time t. Given that, expected steady-state performability can be defined as:

\begin{equation}\label{ssp}
    \centering
      E[X_{SP}] = \sum\limits_{i \in S } w_i * r_i
    \end{equation}

\noindent Expected transient or point performability can be defined as:

\begin{equation}\label{pp}
    \centering
      E[X(t)_{PP}] = \sum\limits_{i \in S } p_i(t) * r_i
    \end{equation}

\noindent \textbf{Markov Reward Modeling for the CDFG:} For a data-flow system, the primary reward associated with each state of the MRM is throughput. For a synchronous data-flow system, the throughput can be evaluated directly from the CDFG of the system. As we consider only non-pipelined systems in this paper, we can define the throughput as the inverse of the number of seconds it takes to execute the CDFG:

\begin{equation}\label{thr1}
    \centering
      Throughput = (1/c_{step}) * (c_{step}/cycle) * (cycles/second)
    \end{equation}

\noindent where, ${c_{step}}$ is the number of control steps in the CDFG. Assuming that each $c_{step}$ takes a single clock cycle and $\eta$ represents the system's clock frequency (clock cycle/second):
\begin{equation}\label{thr2}
    \centering
      Throughput = (1/c_{step}) * \eta
    \end{equation}

%\noindent For the simplicity of the calculation we assumed $\eta = 1$ in our throughput calculation from the CDFG. \\
In our MRM, the operational and degraded states are augmented with associated throughput reward, and all the \emph{failed states} both safe and unsafe ones, are augmented with a throughput reward of zero. The expected throughput (for long run $E[X]$ or for a specific mission time $E[X(t)]$) can be calculated using the equation \ref{ssp} and equation \ref{pp} respecively. In our MRM model, the area that is required, to implement the design on the FPGA, is assumed to be invariant between the states for a specific configuration. The reason is, once the system is implemented on FPGA, the area is fixed (in terms of the total number of LUTs) and if a fault occurs, then the system will be rescheduled or if it fails, then eventually will be scrubbed. So only the control signals will change, not the components. For \emph{overall reward} calculation e.g. to evaluate the throughput-area-reliability trade-offs for a configuration, we use the following equation:

\begin{equation}\label{ovrReward}
    \centering
Overall~reward = (1/A) * E[X]
\end{equation}

\noindent In the above equation, $A$ represents the area of the design and $E[X]$ represents the expected throughput. This equation is similar to \cite{ICYield}, however instead of calculating the reward up to a specified time-step, we use the notion of steady-state throughput ($E[X] = E[X_{SP}]$). Such modeling can be considered as a direct optimization of throughput, area and reliability. Rewards can be weighted based on designer's requirements. For the case study presented in this paper, the rewards are set to equal weight.

\subsection{Characterization Library}
As the SEU rate $\lambda$ is highly dependent on device process technology, architecture, and orbits of interest, so this parameter is different for each device family. We use CREME96 \cite{CREME96} with radiation cross sections from \cite{HQuinn2007} to find per bit upset rate $\lambda_{bit}$ for Xilinx Virtex-5 in the Highly Elliptical Orbit (HEO) and Low Earth Orbit(LEO) orbit. The failure rate for a component can be calculated using the equation as follows:

\begin{equation}\label{failure_rate}
    \centering
\lambda_{component} = \lambda_{bit} \times Number~of~critical~bits
\end{equation}

\begin{table}
\centering
\caption{Characterization library}\label{table_charactariation}

    \begin{tabular}{|c|c|c|c|c|}
    \hline
    \textbf{Component}               & \textbf{No. of} & \textbf{No. of} & \textbf{MTBF} \\
    ~& \textbf{LUTs} & \textbf{essential bits} & \textbf{(days)} \\ \hline
    Wallace Tree Multiplier & 722 & 133503         & 11.85 \\
    Booth Multiplier        & 650 & 130781         & 12.11 \\
    Brant-Kung adder        & 120 & 29675          & 53.36   \\
    Kogge-Stone Adder       & 183 & 41499          & 38.15   \\
    \hline
    \end{tabular}
\end{table}

\noindent For our experiments, $\lambda_{bit} = 7.31 \times 10^{-12}$ SEUs/bit/sec for the HEO orbit.

In order to build a component characterization library that represents the first-order estimation of the SEU effects on the components, we use the \emph{bitgen} feature of Xilinx ISE tool. Using \emph{bitgen}, we identified the \emph{essential bits} which are also known as \emph{potentially~critical~bits}. \emph{Essential bits} are subset of total configuration bits (defined by Xilinx) and they refer to the amount of configuration bits associated with a design mapped in the FPGA. If an essential bit encounters an upset, it changes the design circuitry. However, the upset might not affect the function of the design. In contrast, \emph{critical bits} are defined as those configuration bits that cause a functional failure if they change state. The critical bits are the subset of the essential bits. Note that, it is well known that the number of \emph{critical bits} is less than the number of \emph{potentially critical bits}. More accurate SEU susceptibility analysis can be performed using the fault injection techniques \cite{fault1,SEUFault}, however, for first-order worst-case estimation, it is valid to assume that all the \emph{essential bits} are considered as  \emph{critical bits}. Note that we use the characterization library to obtain the failure rate of the components for the Markov chain model and the methodology is generic enough to be used with a different characterization library with more precise and accurate data, without any major changes.

Table \ref{table_charactariation} presents a first-order worst-case estimate of component failures due to SEUs. We characterize different adder and multiplier components, namely 64-bit Brent-kung adder, 64-bit Kogge-stone adder, 32-bit Wallace-tree multiplier and 32-bit Booth multiplier. The Xilinx Synthesis Technology (XST) tool is used to synthesize the components for Virtex-5 XC5VLX50T device from their HDL codes and the number of required LUTs to implement them (area) is also obtained. We observe that a 32-bit Wallace-tree multiplier has about 0.134 million bits that are sensitive to SEUs. So this multiplier has a worst-case Mean Time Between Failures (MTBF) of 11.85 days for space applications in the HEO orbit. MTBF and $\lambda$ are related to each other using the following equation \cite{MTBF}:

\begin{equation}\label{MTBF}
    \centering
\lambda = {1 \over MTBF}
\end{equation}

\section{Quantitative Analysis using PRISM}

\begin{figure}
\centering
\includegraphics[width=40mm,height=75mm]{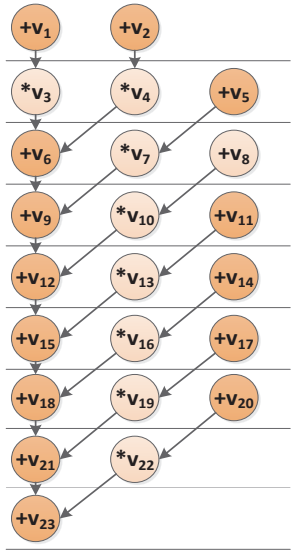}
\caption{CDFG of an FIR filter}\label{fig_fir}
\end{figure}

To illustrate the applicability of the proposed methodology for early design decision, this section presents a Finite Impulse Response (FIR) filter case study from a high-level synthesis benchmark. FIR filters are one of two primary types of digital filters (the other one is Infinite Impulse Response) used in Digital Signal Processing (DSP) applications. FIR filters are commonly used in spacecrafts for noise filtering from images, videos and sensor outputs and spacecraft antennas \cite{FIR-1, FIR-2, FIR-3}.

\figurename~\ref{fig_fir} shows the CDFG for a 16-point FIR Filter \cite{FIR} obtained from \cite{benchmark}. To achieve a schedule with minimum number of control steps, the minimum allocation is two adders and two multipliers for the FIR filter application. At a minimum a pair of one adder and one multiplier is required for successful operation. For our experiments, we consider the 32-bit Kogge-stone adders and 32-bit Wallace tree multipliers as available components from the characterization library, as they require less area (number of LUTs) to implement compared to the others in the characterization library. We must mention that any other adder or multiplier from the component characterization library can be used for the similar analysis. The first part of the case study presents the dependability analysis on different configurations. The latter part of the case study focuses on the performability (throughput with reliability) analysis and overall reward calculation. Overall reward (equation \ref{ovrReward}) gives the expected reward with both area and throughput taken into consideration.

\begin{table}
\centering
\caption{Available design options to evaluate}\label{table_model-generation}

    \begin{tabular}{|c|c|c|c|c|c|}
    \hline
    \textbf{No.}&\textbf{Configuration} & \textbf{Spare} & \textbf{Scrubbing} & \textbf{Rescheduling} \\ \hline
    C1& 2A 2M         & None            & \checkmark                & \checkmark    \\
    C2& 2A 3M         & 1 Mul             & \checkmark                 & \checkmark    \\
    C3& 3A 2M         & 1 Add            & \checkmark                 & \checkmark    \\
    C4& 3A 3M         & 1 Add, 1 Mul            & \checkmark                 & \checkmark    \\
    \hline
    \end{tabular}
\end{table}

\tablename~\ref{table_model-generation} shows the different configurations to evaluate the FIR filter design and its respective fault mitigation strategies.  The first configuration consists of two adders and two multipliers with no redundancy. The second and third configuration consists of one spare multiplier and one spare adder respectively used as redundant components ($cold spare$). Configuration 4 is equipped with full component-level redundancy, with a spare of each type of components. All the four configurations employ scrubbing and rescheduling. In rest of the paper, configurations 1, 2, 3 and 4 will be referred to as C1, C2, C3 and C4 respectively. Also, for brevity when reporting experimental results, the scrub interval and fault detection coverage will be denoted by I (in days) and C, respectively, and their units, when applicable, will be omitted. Before analyzing the model quantitatively, we verify the following LTL style properties (Recall Remark 1) to check the correctness of the model:  \\

%\noindent \emph{Property 1:} $ P>=1~ [ G~ ( (``fail" | ```degrade") \Rightarrow (F~ ``oper") ) ]$ - ``Once the system is in either \emph{degrade} state or \emph{failed} state, it will reach \emph{\emph{operational}} state eventually" with probability 1.\\ \\
\noindent \emph{Correctness Property:} $ filter(forall, P >0~ [  X~oper  ])$ - ``From any reachable state, it is possible to reach the \emph{oper} state in the next step with a probability greater than 0 ".\\

\noindent Note that, as mentioned earlier in the preliminary section, blind scrubbing periodically reconfigures the FPGA, which does not require any fault detection. This sets the requirement (specified as the property above) that the system should be repaired irrespective of its failures, i.e. will be scrubbed even in the \emph{oper} state, which justifies the self-loop in our model. While verifying, PRISM returned \texttt{true}, which means that the correctness property hold in our model.

\begin{table}
\centering
\caption{Configurations vs classes of states}\label{table_config vs class}
    \begin{tabular}{|c|c|c|c|c|c|}
        \hline
        \textbf{Config.} & \textbf{I} & \textbf{Operational} & \textbf{Degraded} & \textbf{Failure} \\
        ~ & \textbf{(days)} & \textbf{(days)} & \textbf{(days)} & \textbf{(days)}\\ \hline

        C1    & 1 & 2989.00     & 609.04   &  51.94  \\
        ~        & 4 & 1937.53     & 1287.04   & 425.42 \\
        ~        & 9 & 1222.40     & 1378.28   & 1049.31 \\
        \hline

        C2    & 1 & 2989.00     & 642.82   &  18.14   \\
        ~        & 4 & 1937.53     & 1492.61   & 219.86  \\
        ~        & 9 & 1222.40     & 1711.59   & 716.00  \\
        \hline
        C3    & 1 & 2989.00     & 613.08   &  47.91   \\
        ~        & 4 & 1937.53     & 1319.58   & 392.88  \\
        ~        & 9 & 1222.40     & 1441.09   & 986.50  \\
        \hline
        C4    & 1 & 2989.00     & 647.06   &  13.93   \\
        ~        & 4 & 1937.53     & 1531.90   & 180.55  \\
        ~        & 9 & 1222.40     & 1795.97   & 631.61  \\
        \hline

    \end{tabular}
\end{table}

In \tablename~\ref{table_config vs class}, using reward-based properties, we analyze the number of days the design spends in different classes of states for a mission time of 10 years and fault coverage C = 0.99, with a value of I = 1, 4 and 9. Different states in the Markov model can be classified into various classes using \emph{formulas} in PRISM language. To calculate the number of days spent in different classes of states, we define a reward structure for each of them. For example, a reward structure \emph{degraded} assigns a state reward of 1 to all states of the model in which the system is in \emph{degraded} mode. A property that can reason about the amount of rewards accumulated over a period of time, is represented using CSL logic in PRISM as follows: \\

\noindent \emph{Property 1:} $R\{``degraded"\} = ? [C<=t]$ - ``the expected cumulative time spent in the degraded mode of the system in the time interval [0, t]".\\

The first column of the table shows the different configurations to be evaluated and the second column shows the associated scrub intervals (I). The  third, fourth, and fifth column presents the number of days the design spends in different classes of states. It is worth mentioning that the fifth column shows the days spent in either \emph{failed safe} or \emph{failed unsafe} states.  All configurations spend approximately similar number of days in \emph{operational state} (rounded to 2 decimal points) for the same scrub intervals. For I = 9, configuration C1 that has no redundant components shows the worst result. Interestingly, we observe that adding an extra adder as spare does not help much whereas adding an extra multiplier as spare significantly reduces the number of days spent in \emph{failed} states. In configuration C4, the added spares for both adder and multiplier provide the best result in terms of availability. This is obvious but will cost more area on the FPGA. Configuration C1 spends the least number of days and configuration C4 spends the highest number of days in \emph{degraded} states. For many safety-critical applications, low performance for a period of time is acceptable. For such systems the number of days spent in \emph{failed} states is a major concern and hence, configuration C4 and configuration C2 are the two best candidates.

\begin{figure}[!t]
\centering
\includegraphics[width= \textwidth]{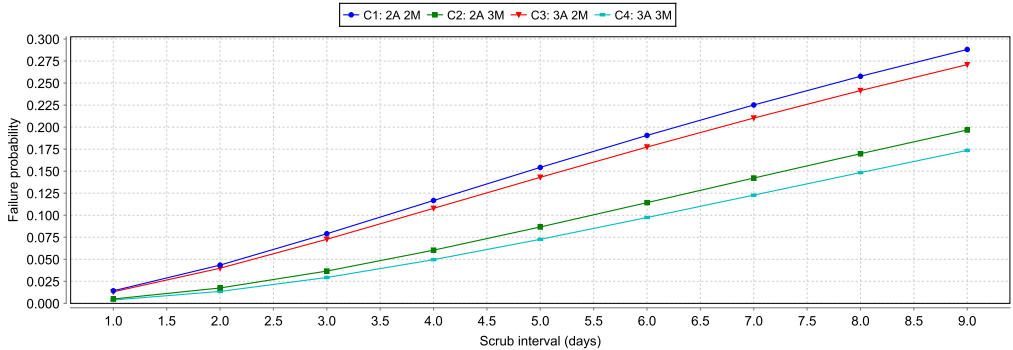}
\caption{Failure probability vs I (scrub interval)} \label{fig_config}
\end{figure}

Steady state analysis of a design is useful to evaluate its dependability in the long-run. In \figurename~\ref{fig_config}, we calculate the steady-state failure probability (safe or unsafe) and compare the results of the four available configurations, with respect to different scrub intervals (I is varied from 1 to 9) and same coverage (C = 0.99). The steady-sate failure probability for a given configuration can be analyzed in PRISM using the following property: \\

\noindent \emph{Property 2:} $S = ? ~[~failed_{safe}~ + ~failed_{unsafe}~]$ - ``the long-run non-availability of the system". \\

\noindent The experimental results show that for configuration C1, the failure probability varies from 0.014 to 0.288 depending on the value of I. Configuration C2 has a lower failure probability than configuration C3 for all the scrub intervals. The failure probability of configuration C4 for all different scrub rates shows the best result with associated extra area overhead. From the results, we observe that configuration C2 is really an attractive alternative to configuration C4 (even for I = 7, the probability varies by only 0.023). On the other hand, configuration C1 and configuration C3 offer similar results over the long-run. Another conclusion that can be added is, for a value of $I > 2$ , the failure probability increases for all the configurations. For a value of $I \leq 2$, configuration C1 and configuration C3, and, configuration C2 and configuration C4 has almost same failure probability.

\begin{figure}[!t]
\centering
\includegraphics[width=\textwidth]{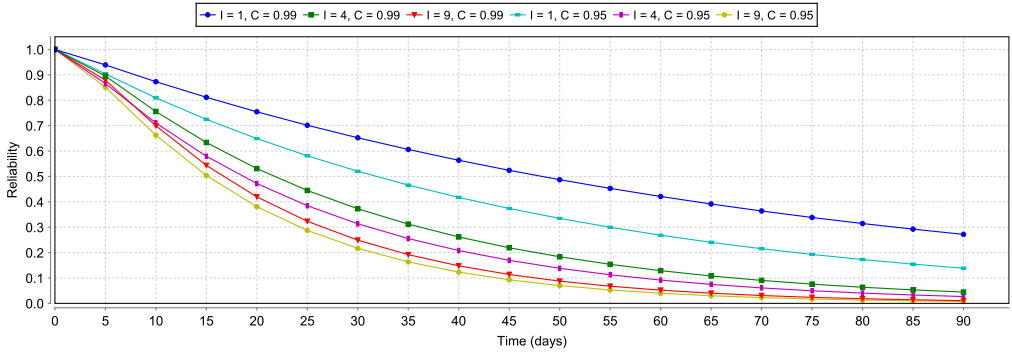}
\caption{Reliability vs I (scrub interval)} \label{fig_reliability_cov}
\end{figure}

\begin{figure}
\centering
\includegraphics[width=\textwidth]{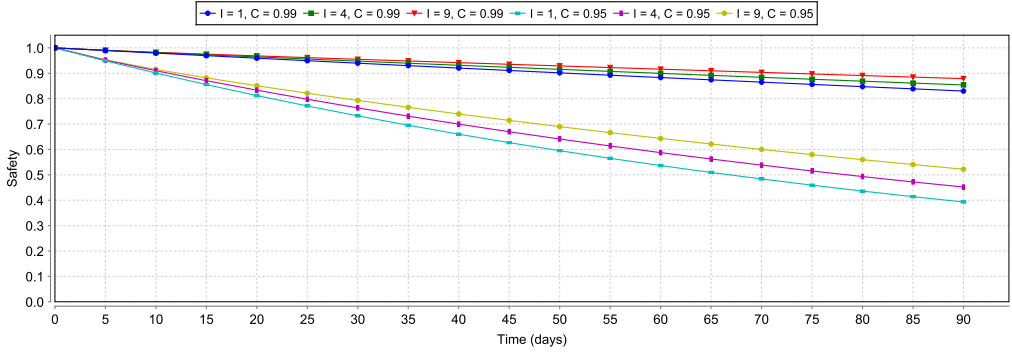}
\caption{Safety vs scrub I (scrub interval)} \label{fig_safety_cov}
\end{figure}

\figurename~\ref{fig_reliability_cov} and \figurename~\ref{fig_safety_cov} shows the effect of coverage \emph{C} on reliability and safety respectively, for different values of I, for a mission time (T) of maximum 3 months (T is varied from 1 to 90 days). For this part of the experiment (reliability and safety analysis), we consider configuration C1 with two adders and two multipliers, however any other configuration can also be analyzed in the similar fashion. The properties used to analyze reliability and safety in PRISM are as follows:\\

\noindent \emph{Property~3 (Safety)} : $P~=~?~ [~ G~[0,~T] ~operational~|~degraded~|\\~failed_{safe}~ ]$ - ``The probability that the system will be either in a \emph{operational}, \emph{degraded}, or \emph{failed safe} state in first T days". \\

\noindent \emph{Property~4 (Reliability)} : $~P~=~?~ [~ G~[0,~T]~ operational~|~\\degraded~ ]$ - ``The probability that the system will be either in a \emph{operational} or \emph{degraded} state in first T days". \\

\noindent \figurename~\ref{fig_reliability_cov} shows some interesting results for reliability evaluation. Configuration C1 has the highest reliability for I = 1 and C = 0.99. We observe that, with the same coverage, for a delayed scrub of I = 4, configuration C1 has lower reliability than the same configuration with I = 1 and C = 0.95. So, a high coverage does not by itself guarantee a high reliability, particularly if the scrub interval is long. In contrast, if the scrub interval is fixed, then increasing the coverage will always increase the reliability. For example, the design with I = 4 and C = 0.99 has a higher reliability compared with the design with I = 4 and C = 0.95. In \figurename~\ref{fig_safety_cov}, we observe that, for C=0.99, the safety of the system never goes below 0.83 in the first 3 months, even for the most delayed scrub interval (I = 9). When we analyze the system for C = 0.95, it shows how drastically the safety of the system falls. For C = 0.95, the safety of the system drops up to 0.39 for a mission time of 3 months. It is also noticeable that if the value of $I$ increases, then the distance between the safety values also get wider even for the same coverage.

\begin{figure}
\centering
\includegraphics[width=\textwidth]{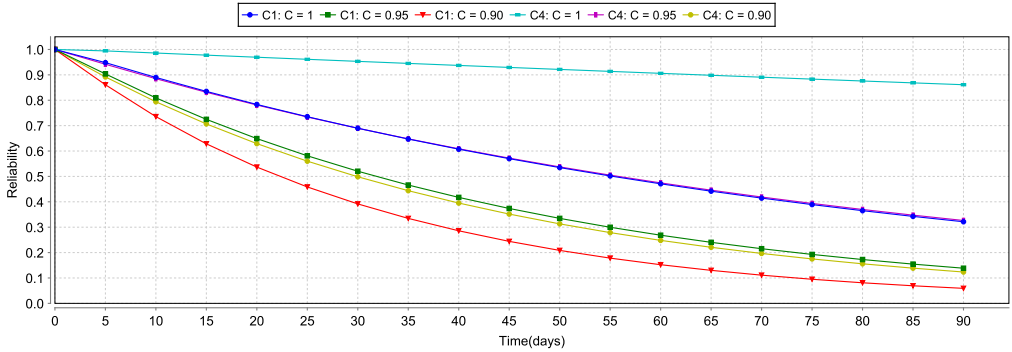}
\caption{Impact of C (coverage) on the design with/without redundancy for I = 1)} \label{fig:relvscov}
\end{figure}

\begin{figure}
\centering
\includegraphics[width=\textwidth]{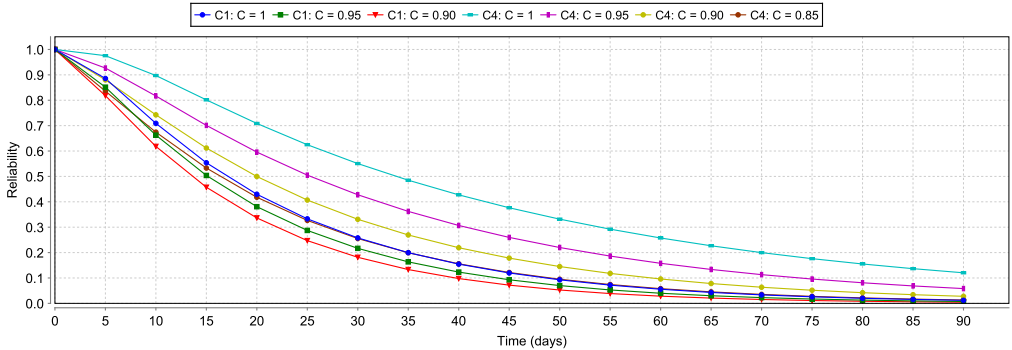}
\caption{Impact of C (coverage) on the design with/without redundancy for I = 9)} \label{fig:relvscovvsscrb}
\end{figure}

\noindent \figurename~\ref{fig:relvscov} reveals an important observation to compare the available design options. We compare configuration C1 with no redundancy and configuration C4 with full redundancy for three different values of coverage C and I = 1 (since the model is parametric, any other parameter combinations can also be easily evaluated). We observe that, for perfect coverage (C = 1), indeed the configuration with redundancy gives better reliability. However, for lower coverage values, such as C = 0.95, configuration C4 with redundancy gives almost the same reliability compared to the configuration C1 with no redundancy with perfect coverage. For even lower coverage value, redundancy fails to improve the reliability compared to the configuration C1 for the cases where it has better coverage. This experiment shows that a design option with redundancy is not always the best choice with lower coverage. For instance, all C4 curves for which $C < 0.95$ produce a reliability less than C1 with C = 1. We redo this experiment for a delayed scrub, I = 9 and plot the results in \figurename~\ref{fig:relvscovvsscrb}. This experiment shows that if a design option has coverage more that 0.85, then the design option with redundancy provides a better reliability. From this, we can conclude that, if a system employs longer scrub interval, then a design option with redundancy can provide better reliability even with lower coverage, compared with the design option with no redundancy, with same coverage. However, if the coverage goes lower beyond a certain point, indeed redundancy will not help improving the reliability. Comparison of \figurename~\ref{fig:relvscov} and \figurename~\ref{fig:relvscovvsscrb} also indicates that redundancy is more useful for improving reliability in the cases where scrub interval is longer. For systems with fast scrubbing capability, rescheduling can be a good alternative to redundancy-based solutions.

\begin{table}
\centering
%\begin{center}
\caption{Overall reward calculation}
\label{table_areathroughputreward}
\begin{tabular}{|c|c|c|c|c|c|}
        \hline
        \textbf{I} & \textbf{Config.} & \textbf{Normalized} & \textbf{Area}  & \textbf{Norm.} & \textbf{Overall}  \\
        \textbf{(days)} & ~ &  \textbf{Expected} &  \textbf{(No. of} & \textbf{Area} &  \textbf{Reward} \\
        ~ & ~ & \textbf{Throughput} & \textbf{LUTs)}& ~ & ~ \\ \hline
        1 & C1          & 0.955                & 1810                  & 0.667 & 1.432            \\
        ~ & C2          & 0.974                & 2532                  & 0.932 & 1.045            \\
        ~ & C3          & 0.973                & 1934                  & 0.734 & 1.326            \\
        ~ & C4          & 0.993                & 2765                  & 1.000 & 0.993            \\
        \hline

        4 & C1          & 0.811                & 1810                  & 0.667 & 1.216            \\
        ~ & C2          & 0.876                & 2532                  & 0.932 & 0.940            \\
        ~ & C3          & 0.856                & 1934                  & 0.734 & 1.166            \\
        ~ & C4          & 0.931                & 2765                  & 1.000 & 0.931            \\
        \hline

        9 & C1          & 0.628                & 1810                  & 0.667 & 0.942            \\
        ~ & C2          & 0.717                & 2532                  & 0.932 & 0.769            \\
        ~ & C3          & 0.684                & 1934                  & 0.734 & 0.932            \\
        ~ & C4          & 0.790                & 2765                  & 1.000 & 0.790            \\
        \hline

\end{tabular}
%\end{center}
\end{table}
%\end{center}

For performability and throughput-area analysis, table \ref{table_areathroughputreward} shows the expected throughput and long-run \emph{overall reward} calculation for various scrub intervals with C = 0.99. The rewards are setup so that the area and expected throughput have equal weights. Both the area and throughput were normalized between 0 and 1 in order to not skew the reward numbers. For every configuration, the maximum throughput (throughput in the initial state) is used to normalize the throughput for other states in the Markov reward model. Similarly, the maximum area is used to normalize the other area values among different configurations. In our model, a reward structure \emph{throughput} assigns a normalized throughput reward to all the operational or degraded states. All the \emph{failed safe }and \emph{failed unsafe} states are augmented with a throughput reward of zero. Steady-state expected throughput (normalized) for a configuration can be analyzed in PRISM using the property as follows and shown in column 3: \\

\noindent \emph{Property 5:} $ R~ \{``Expected ~throughput"\} ~=~ ? ~ [~S~] $ - ``The expected throughput of the system". \\

\begin{figure}[!t]
\centering
\includegraphics[width=\textwidth]{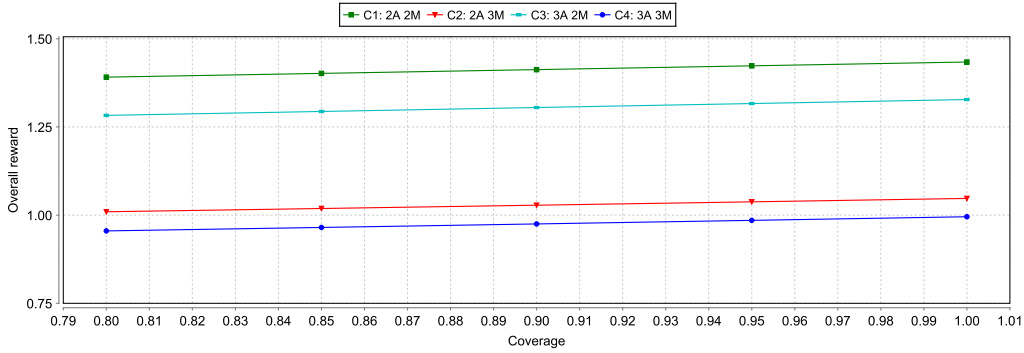}
\caption{Impact of C (coverage) on for performability-area trade-off evaluation for I = 1} \label{fig:ovreward}
\end{figure}

\begin{figure}[!t]
\centering
\includegraphics[width=\textwidth]{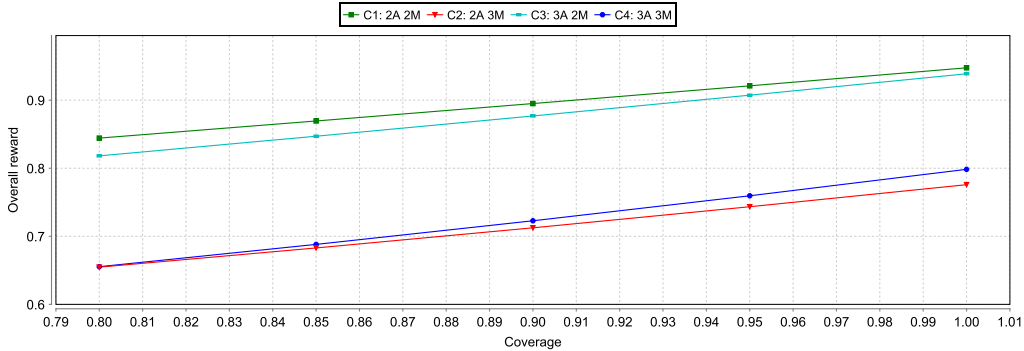}
\caption{Impact of C (coverage) on for performability-area trade-off evaluation for I = 9} \label{fig:ovreward-1}
\end{figure}

\noindent Column 4 shows the area of each configuration and their normalized value is shown in column 5. Column 6 shows the overall area-throughput reward (overall reward) for each configuration. The reward for each configuration is calculated by multiplying the value of column 3 with the reciprocal of the normalized area. Based on the equal reward weighting, configuration C1 which has no redundancy (spare components), shows the best throughput-area reward for all the values of I. This indicates that the extra reliability provided by the redundancy is not always useful to suppress the extra area overhead. However, rescheduling with scrubbing is good enough to serve as a fault recovery and repair mechanism in such cases. Another important observation is that adding a spare adder significantly improves the throughput-area reward, much more than adding a spare multiplier. If performance is the main concern of the design, then the expected throughput results from column 3 suggests configuration C4 as the best choice to implement. It clearly shows, how the inclusion of throughput-area metrics can influence design decisions toward solutions that differs from those resulting from an analysis based on either dependability (as in Table \ref{table_config vs class}) or performability metric alone. Such an analysis, using the proposed methodology, can be very useful at early design stages for designers of safety-critical applications concerned with dependability, performance and area constraints.

To analyze the impact of coverage on performability-area trade-off, we evaluate property 5 for scrub interval I = 1 and show the result in \figurename~\ref{fig:ovreward}. We find that, from lower to higher coverage, the trend is the same, configuration C1 with no redundancy keep dominating the overall reward graph. This supports the conclusion derived from Table~\ref{table_areathroughputreward} that redundancy is not always useful to suppress the extra area overhead for all coverage points. In contrast, when we redo this experiment for a comparatively delayed scrub I = 9, we clearly notice the relationship between $I$ and $C$ reflected on the overall reward as shown in \figurename~\ref{fig:ovreward-1}. The configuration C1 with no redundancy is still dominating, but the rewards accumulated by configuration C3 approaches closer to configuration C1 with increasing values of $C$. In contrast, for lower coverage values, configuration C2 and configuration C4 accumulate almost similar reward, however the gap between them expands with increasing values of $C$. Such phenomena was not observed in \figurename~\ref{fig:ovreward}, but in \figurename~\ref{fig:ovreward-1} it is visible for delayed scrub interval.

\section{Conclusion \& Future Works}
This paper illustrated how probabilistic model checking, a formal verification technique which has already been applied to a wide range of domains, can be used to analyze designs  at early stage for space applications. The design options are modeled using a Markov reward model, that captures the possible failures, fault detection coverage and repairs possible in high-altitude radiation environment. Afterwards, a wide range of properties are exhaustively and automatically verified to evaluate the design options, in terms of throughput, area and dependability. Such analysis is useful to reduce the overall design cost and effort. Quantitative results from an FIR filter case study demonstrated how the proposed methodology can be applied to drive the design process. The PRISM model checker includes multiple model checking engines, the majority of which are based on symbolic implementations (using binary decision diagrams and their extensions). These engines enables the probabilistic verification of models of up to $10^{10}$ states (on average, PRISM handles models with up to $10^7 - 10^8$ states). PRISM also features a variety of advanced techniques such as abstraction refinement and symmetry reduction. It is worth mentioning that it also supports approximate/statistical model checking through a discrete event simulation engine. So considering the capability of PRISM model checker, it is possible to analyze large systems using our methodology. However, since the CDFG rescheduling part of the methodology is not fully automated yet, this currently restricts us from doing so. Future works include the automation of this process to generate the PRISM code for a given configuration automatically, and to analyze designs in the presence of other kinds of possible failures due to SEUs, such as aging, electromigration, hot electron effects, Negative-Bias Temperature Instability (NBTI) and Single-Event Functional Interrupts (SEFIs).

% conference papers do not normally have an appendix

% use section* for acknowledgement
\section*{Acknowledgments}
This research work is a part of the AVIO-403 project financially supported by the Consortium for Research and Innovation in Aerospace in Quebec (CRIAQ), Fonds de Recherche du Qu\'ebec - Nature et Technologies (FRQNT) and the Natural Sciences and Engineering Research Council of Canada (NSERC). The authors would also like to thank Bombardier Aerospace, MDA Space Missions and the Canadian Space Agency (CSA) for their technical guidance and financial support.

%% If you have bibdatabase file and want bibtex to generate the
%% bibitems, please use
%%
%\section*{References}

\bibliographystyle{elsarticle-num}
\bibliography{myBib}

\begin{thebibliography}{10}
\expandafter\ifx\csname url\endcsname\relax
  \def\url#1{\texttt{#1}}\fi
\expandafter\ifx\csname urlprefix\endcsname\relax\def\urlprefix{URL }\fi
\expandafter\ifx\csname href\endcsname\relax
  \def\href#1#2{#2} \def\path#1{#1}\fi

\bibitem{XilinxRosetta}
A.~Lesea, Continuing experiments of atmospheric neutron effects on deep
  submicron integrated circuits ({WP286} v1.1) (October 2011).

\bibitem{NASAScrub}
P.~Adell, G.~Allen, G.~Swift, S.~McClure, Assessing and mitigating radiation
  effects in {X}ilinx {SRAM} {FPGA}s, in: Radiation and Its Effects on
  Components and Systems (RADECS), 2008 European Conference on, 2008, pp.
  418--424.

\bibitem{Xilinx5QV}
G.~Swift, C.~Carmichael, G.~Allen, G.~Madias, E.~Miller, R.~Monreal, Compendium
  of {XRTC} radiation results on all single-event effects observed in the
  {Virtex-5QV}, ReSpace/MAPLD.

\bibitem{XilinxTMR}
C.~Carmichael, Triple module redundancy design techniques for virtex {FPGA}s
  ({XAPP}197 v1.0.1), {X}ilinx corporation, 2006.

\bibitem{ScrubPower}
G.~Nazar, L.~Santos, L.~Carro, Scrubbing unit repositioning for fast error
  repair in fpgas, in: Compilers, Architecture and Synthesis for Embedded
  Systems (CASES), 2013 International Conference on, 2013, pp. 1--10.
\newblock \href {http://dx.doi.org/10.1109/CASES.2013.6662506}
  {\path{doi:10.1109/CASES.2013.6662506}}.

\bibitem{scrub_power_thesis2_dynamic}
D.~Llamocca, {Dynamically reconfigurable management of energy, performance, and
  accuracy applied to digital signal, image, and video Processing Applications,
  Ph.D. thesis, The University of New Mexico, USA, 2012}.

\bibitem{scrub_power_thesis}
J.~D. Snodgrass, {Low-power fault tolerance for spacecraft FPGA-based numerical
  computing, Ph.D. thesis, Naval Postgraduate School, Monterey, California,
  USA, 2006 }.

\bibitem{voyager}
J.~Warwick, J.~Pearce, A.~Riddle, J.~Alexander, M.~Desch, M.~Kaiser,
  J.~Thieman, T.~Carr, S.~Gulkis, A.~Boischot, C.~C. HARVEY, B.~M. PEDERSEN,
  Voyager 1 planetary radio astronomy observations near jupiter, Science
  204~(4396) (1979) 995--998.

\bibitem{khaza}
K.~A. Hoque, O.~Ait~Mohamed, Y.~Savaria, C.~Thibeault, {Early Analysis of Soft
  Error Effects for Aerospace Applications Using Probabilistic Model Checking},
  in: C.~Artho, P.~C. Ölveczky (Eds.), Formal Techniques for Safety-Critical
  Systems, Vol. 419 of Communications in Computer and Information Science,
  Springer International Publishing, 2014, pp. 54--70.

\bibitem{cov1}
S.~Blanc, P.~Gil, Improving the multiple errors detection coverage in
  distributed embedded systems, in: Reliable Distributed Systems, 2003.
  Proceedings. 22nd International Symposium on, 2003, pp. 303--312.
\newblock \href {http://dx.doi.org/10.1109/RELDIS.2003.1238081}
  {\path{doi:10.1109/RELDIS.2003.1238081}}.

\bibitem{cov2}
J.~Bechta~Dugan, K.~Trivedi, Coverage modeling for dependability analysis of
  fault-tolerant systems, Computers, IEEE Transactions on 38~(6) (1989)
  775--787.
\newblock \href {http://dx.doi.org/10.1109/12.24286}
  {\path{doi:10.1109/12.24286}}.

\bibitem{cov3}
M.~Cukier, D.~Powell, J.~Ariat, Coverage estimation methods for stratified
  fault-injection, Computers, IEEE Transactions on 48~(7) (1999) 707--723.
\newblock \href {http://dx.doi.org/10.1109/12.780878}
  {\path{doi:10.1109/12.780878}}.

\bibitem{rescheduling}
B.~R. Borgerson, R.~F. Freitas, A reliability model for gracefully degrading
  and standby-sparing systems, IEEE Transaction on Computers 24~(5) (1975)
  517--525.

\bibitem{CDFGFault}
I.~Hong, M.~Potkonjak, R.~Karri, Heterogeneous {BISR}-approach using system
  level synthesis flexibility, in: Design Automation Conference 1998.
  Proceedings of the ASP-DAC '98. Asia and South Pacific, 1998, pp. 289--294.

\bibitem{PMC1stochastic}
M.~Kwiatkowska, G.~Norman, D.~Parker, Stochastic model checking, in: Formal
  methods for performance evaluation, Springer, 2007, pp. 220--270.

\bibitem{advancesPMC}
M.~Kwiatkowska, G.~Norman, D.~Parker, Advances and challenges of probabilistic
  model checking, in: Communication, Control, and Computing (Allerton), 2010
  48th Annual Allerton Conference on, IEEE, 2010, pp. 1691--1698.

\bibitem{MRM2}
W.~J. Stewart, Introduction to the numerical solution of Markov Chains,
  Princeton University Press, 1994.

\bibitem{intro-related}
J.~Kastil, M.~Straka, L.~Miculka, Z.~Kotasek, Dependability analysis of fault
  tolerant systems based on partial dynamic reconfiguration implemented into
  {FPGA}, in: Digital System Design (DSD), 2012 15th Euromicro Conference on,
  IEEE, 2012, pp. 250--257.

\bibitem{soap}
Q.~Martin, A.~D. George, Scrubbing optimization via availability prediction
  {(SOAP)} for reconfigurable space computing, in: High Performance Extreme
  Computing (HPEC), 2012 IEEE Conference on, IEEE, 2012, pp. 1--6.

\bibitem{Related_DFG_MRM}
V.~V. Kumar, R.~Verma, J.~Lach, J.~Bechta~Dugan, A markov reward model for
  reliable synchronous dataflow system design, in: Dependable Systems and
  Networks, 2004 International Conference on, 2004, pp. 817--825.

\bibitem{website:ISOGraph}
ISOGraph, \url{http://www.isograph-software.com}.

\bibitem{PMC2prism}
M.~Kwiatkowska, G.~Norman, D.~Parker, Prism 4.0: Verification of probabilistic
  real-time systems, in: Computer aided verification, Springer, 2011, pp.
  585--591.

\bibitem{intvsExtScrub}
M.~Berg, C.~Poivey, D.~Petrick, D.~Espinosa, A.~Lesea, K.~LaBel, M.~Friendlich,
  H.~Kim, A.~Phan, Effectiveness of internal versus external {SEU} scrubbing
  mitigation strategies in a {X}ilinx {FPGA}: Design, test, and analysis,
  Nuclear Science, IEEE Transactions on 55~(4) (2008) 2259--2266.

\bibitem{Khaza_MEMOCODE2014}
K.~A. Hoque, O.~A. Mohamed, Y.~Savaria, C.~Thibeault, {Probabilistic Model
  Checking Based DAL Analysis to Optimize a Combined TMR-Blind-Scrubbing
  Mitigation Technique for FPGA-Based Aerospace Applications}, in:
  International Conference on Formal Methods and Models for Co-Design,
  ACM-IEEE, 2014.

\bibitem{Clarke86automaticverification}
E.~M. Clarke, E.~A. Emerson, A.~P. Sistla, Automatic verification of
  finite-state concurrent systems using temporal logic specifications, ACM
  Transactions on Programming Languages and Systems 8 (1986) 244--263.

\bibitem{pnueli1977temporal}
A.~Pnueli, The temporal logic of programs, in: Foundations of Computer Science,
  1977., 18th Annual Symposium on, IEEE, 1977, pp. 46--57.

\bibitem{Baier99approximatesymbolic}
C.~Baier, J.-P. Katoen, H.~Hermanns, Approximate symbolic model checking of
  continuous-time markov chains (extended abstract) (1999).

\bibitem{kwiatkowska2006model}
M.~Kwiatkowska, G.~Norman, A.~Pacheco, Model checking expected time and
  expected reward formulae with random time bounds, Computers \& Mathematics
  with Applications 51~(2) (2006) 305--316.

\bibitem{latency1}
N.~Coste, H.~Hermanns, E.~Lantreibecq, W.~Serwe, Towards performance prediction
  of compositional models in industrial gals designs, in: Computer Aided
  Verification, Springer, 2009, pp. 204--218.

\bibitem{latency2}
P.~A. Beerel, R.~O. Ozdag, M.~Ferretti, A designer's guide to asynchronous
  VLSI, Cambridge University Press, 2010.

\bibitem{FIR}
R.~Karri, A.~Orailoglu, High-level synthesis of fault-secure
  microarchitectures, in: Design Automation, 1993. 30th Conference on, 1993,
  pp. 429--433.

\bibitem{TosunHLS}
S.~Tosun, N.~Mansouri, E.~Arvas, Y.~Xie, Reliability-{C}entric {H}igh-{L}evel
  {S}ynthesis, in: {Proceedings of Desing Automation and Test in Europe
  (DATE)}, 2005.

\bibitem{related_MRM1}
M.~D. Beaudry, Performance-related reliability measures for computing systems,
  Computers, IEEE Transactions on C-27~(6) (1978) 540--547.

\bibitem{related_MRM2}
R.~Huslende, A combined evaluation of performance and reliability for
  degradable systems, in: Proceedings of the 1981 ACM SIGMETRICS conference on
  Measurement and modeling of computer systems, ACM, 1981, pp. 157--164.

\bibitem{ana1}
E.~Cheshmikhani, H.~R. Zarandi, Probabilistic analysis of dynamic and temporal
  fault trees using accurate stochastic logic gates, Microelectronics
  Reliability.

\bibitem{covrel-1}
L.~Xing, S.~V. Amari, C.~Wang, Reliability of k-out-of-n systems with
  phased-mission requirements and imperfect fault coverage, Reliability
  Engineering \& System Safety 103 (2012) 45--50.

\bibitem{covrel-2}
T.~DeLong, D.~Smith, B.~Johnson, Dependability metrics to assess
  safety-critical systems, Reliability, IEEE Transactions on 54~(3) (2005)
  498--505.
\newblock \href {http://dx.doi.org/10.1109/TR.2005.853567}
  {\path{doi:10.1109/TR.2005.853567}}.

\bibitem{covrel-3}
S.~Verlinden, G.~Deconinck, B.~Coup{\'e}, Hybrid reliability model for nuclear
  reactor safety system, Reliability Engineering \& System Safety 101 (2012)
  35--47.

\bibitem{related_casestudy}
R.~M. Smith, K.~Trivedi, A.~Ramesh, Performability analysis: measures, an
  algorithm, and a case study, Computers, IEEE Transactions on 37~(4) (1988)
  406--417.

\bibitem{LibraryBasedSER}
C.~Thibeault, Y.~Hariri, S.~R. Hasan, C.~Hobeika, Y.~Savaria, Y.~Audet, F.~Z.
  Tazi, A library-based early soft error sensitivity analysis technique for
  {SRAM}-based {FPGA} design, J. Electronic Testing 29~(4) (2013) 457--471.

\bibitem{GAUT}
P.~Coussy, C.~Chavet, P.~Bomel, D.~Heller, E.~Senn, E.~Martin, {GAUT}: A
  high-level synthesis tool for {DSP} applications, in: P.~Coussy, A.~Morawiec
  (Eds.), High-Level Synthesis, Springer Netherlands, 2008, pp. 147--169.

\bibitem{Paulin}
P.~G. Paulin, J.~P. Knight, Force-directed scheduling for the behavioral
  synthesis of asics, Computer-Aided Design of Integrated Circuits and Systems,
  IEEE Transactions on 8~(6) (1989) 661--679.

\bibitem{li2013reliability1}
Y.~Li, B.~Nelson, M.~Wirthlin, {Reliability models for SEC/DED memory with
  scrubbing in FPGA-based designs}, IEEE Transactions on Nuclear Science 60~(4)
  (2013) 2720--2727.

\bibitem{li2013reliability2}
Y.~Li, {Reliability Techniques for Data Communication and Storage in FPGA-Based
  Circuits}.

\bibitem{li2013reliability3}
D.~Mirzoyan, {Fault-tolerant memories in FPGA based embedded systems},
  Citeseer, 2009.

\bibitem{Erlang}
A.~David, S.~Larry, {The least variable phase type distribution is Erlang},
  Stochastic Models 3~(3) (1987) 467--473.

\bibitem{Hoque_DATE2015}
K.~A. Hoque, O.~Mohamed, Y.~Savaria, Towards an accurate reliability,
  availability and maintainability analysis approach for satellite systems
  based on probabilistic model checking, in: Design, Automation Test in Europe
  Conference Exhibition (DATE), 2015, 2015, pp. 1635--1640.

\bibitem{hoque2016early}
K.~A. Hoque, {Early Dependability Analysis of FPGA-Based Space Applications
  Using Formal Verification}, Ph.D. thesis, Concordia University (2016).

\bibitem{hoque2016applying}
K.~A. Hoque, O.~A. Mohamed, Y.~Savaria, {Applying Formal Verification to Early
  Assessment of FPGA-based Aerospace Applications: Methodology and Experience},
  in: 10th IEEE International Systems Conference, IEEE, 2016.

\bibitem{dubrova2013fault}
E.~Dubrova, Fault-tolerant design, Springer, 2013.

\bibitem{rennels1984fault}
D.~A. Rennels, Fault-tolerant computing --- concepts and examples, IEEE
  Transactions on computers 100~(12) (1984) 1116--1129.

\bibitem{stott2008fault}
E.~Stott, P.~Sedcole, P.~Y. Cheung, Fault tolerant methods for reliability in
  fpgas, in: 2008 International Conference on Field Programmable Logic and
  Applications, IEEE, 2008, pp. 415--420.

\bibitem{stott2008fault1}
C.~R. Elks, M.~Reynolds, N.~George, M.~Miklo, S.~Bingham, R.~Williams, B.~W.
  Johnson, M.~Waterman, J.~Dion, Application of a fault injection based
  dependability assessment process to a commercial safety critical nuclear
  reactor protection system, in: 2010 IEEE/IFIP International Conference on
  Dependable Systems \& Networks (DSN), IEEE, 2010, pp. 425--430.

\bibitem{performance_MRM}
R.~A. Sahner, K.~Trivedi, A.~Puliafito, Performance and reliability analysis of
  computer systems: an example-based approach using the sharpe software
  package.

\bibitem{ICYield}
V.~Kumar, J.~Lach, {IC} modeling for yield-aware design with variable defect
  rates, in: Reliability and Maintainability Symposium, 2005. Proceedings.
  Annual, 2005, pp. 489--495.

\bibitem{CREME96}
A.~Tylka, J.~Adams, P.~Boberg, B.~Brownstein, W.~Dietrich, E.~Flueckiger,
  E.~Petersen, M.~Shea, D.~Smart, E.~Smith, {CREME96}: A revision of the cosmic
  ray effects on micro-electronics code, Nuclear Science, IEEE Transactions on
  44~(6) (1997) 2150--2160.

\bibitem{HQuinn2007}
H.~Quinn, K.~Morgan, P.~Graham, J.~Krone, M.~Caffrey, Static proton and heavy
  ion testing of the {X}ilinx {V}irtex-5 device, in: Radiation Effects Data
  Workshop, 2007 IEEE, Vol.~0, 2007, pp. 177--184.

\bibitem{fault1}
W.~Mansour, R.~Velazco, {SEU} fault-injection in {VHDL}-based processors: A
  case study, J. Electronic Testing 29~(1) (2013) 87--94.

\bibitem{SEUFault}
P.~Kenterlis, N.~Kranitis, A.~M. Paschalis, D.~Gizopoulos, M.~Psarakis, A
  low-cost {SEU} fault emulation platform for {SRAM}-based {FPGA}s, in: IOLTS,
  2006, pp. 235--241.

\bibitem{MTBF}
A.~F. Van~Putten, Electronic measurement systems: theory and practice, CRC
  Press, 1996.

\bibitem{FIR-1}
S.~Visser, A.~Dawood, J.~Williams, {FPGA based real-time adaptive filtering for
  space applications}, in: Field-Programmable Technology, 2002. (FPT).
  Proceedings. 2002 IEEE International Conference on, 2002, pp. 322--326.

\bibitem{FIR-2}
T.~Fry, S.~Hauck, Hyperspectral image compression on reconfigurable platforms,
  in: Field-Programmable Custom Computing Machines, 2002. Proceedings. 10th
  Annual IEEE Symposium on, 2002, pp. 251--260.

\bibitem{FIR-3}
T.~M. Braun, Satellite Communications payload and system, John Wiley \& Sons,
  2012.

\bibitem{benchmark}
S.~P. Mohanty, N.~Ranganathan, E.~Kougianos, P.~Patra, Low-power high-level
  synthesis for nanoscale CMOS circuits, Springer Science \& Business Media,
  2008.

\end{thebibliography}

%% else use the following coding to input the bibitems directly in the
%% TeX file.

%\begin{thebibliography}{00}
%
%%% \bibitem{label}
%%% Text of bibliographic item
%
%\bibitem{}

%\end{thebibliography}
\end{document}